\documentclass[pra,twocolumn,longbibliography,nofootinbib,floatfix]{revtex4-1}
\usepackage[utf8]{inputenc}
\usepackage{bm,amsfonts,amsmath,amssymb}
\usepackage{dcolumn,xfrac}
\usepackage{graphicx}
\usepackage{natbib}
\usepackage{braket}
\usepackage{tipa}
\usepackage{booktabs}
\usepackage{multirow}
\usepackage{cases}
\usepackage{dcolumn}

\def\veps{\varepsilon}

\usepackage{mathtools}

\begin{document}
\title{Basis set calculations of heavy atoms}
\author{M. G. Kozlov$^{1,2}$}
\author{Yu. A. Demidov$^{1,2}$}
\author{M. Y. Kaygorodov$^{1}$}
\author{E. V. Triapitsyna$^{1}$}

\affiliation{$^1$ Petersburg Nuclear Physics Institute NRC ``Kurchatov Institute'', Gatchina 188300, Russia \\
$^2$ St.~Petersburg Electrotechnical University ``LETI'', St. Petersburg 197376, Russia}

\begin{abstract}
Most modern calculations of many-electron atoms use basis sets of atomic orbitals. An accurate account for the electronic correlations in heavy atoms is very difficult computational problem and optimization of the basis sets can reduce computational costs and increase final accuracy. Here we suggest a simple differential ansatz to form virtual orbitals from the Dirac-Fock orbitals of the core and valence electrons. We use basis sets with such orbitals to calculate different properties in Cs including hyperfine structure constants and QED corrections to the valence energies and to the $E1$ transition amplitudes.
\end{abstract}
\maketitle
\section{Introduction}\label{Sec:Intro}

Many efficient methods of atomic calculations, such as configuration interaction (CI), many-body perturbation theory (MBPT), coupled cluster (CC), or their combinations, employ basis sets of one-electron orbitals. 
For example, the methods CI+MBPT and CI+CC \cite{DFK96,Koz04,Dzu05a,SKJJ09,KaBe19} use CI for valence electrons, where the electronic-correlations effects are strong, and account for weaker core-valence correlations by means of either the second order MBPT, or (linearized) CC method. 
In these methods, the relativistic effects are treated within the no-pair approximation for the Dirac-Coulomb-Breit Hamiltonian.
On the other hand, QED effects may be approximately accounted for using the model-QED-potential approach \cite{2003_PyykkoP_JPB36,FlaGin05,TuBe13}.
For heavy polyvalent atoms such calculations become very computationally expensive \cite{Cheung2021}.
That is why it is very important to develop efficient basis sets, which provide high accuracy at reasonable length.
\par
The Dirac-Fock (DF) method serves as initial approximation for most of the atomic calculations. 
However in practice, it is useful to keep DF orbitals in the basis set without re-expanding them in some nonphysical basic orbitals.
On the other hand, the highly excited DF orbitals are usually ineffective in accounting for the correlations between valence electrons, as their radius grows too rapidly. 
Because of that the B-splines \cite{JBS88}, Sturm orbitals \cite{TL03}, or other simple orbitals \cite{Bv83,KPF96} usually turn out to be more useful. 
Therefore, an efficient basis set has to include different types of orbitals. 
An effective method to merge two subsets of orbitals in one basis set was suggested in Ref.~\cite{KozTup19}. 
The first subset consisted of DF orbitals for core and valence shells and the second subset consisted of B-splines, whose parameters were optimized to complement the first subset.  
Such mixed basis sets were tested for calculations of the valence energies of Au and Fr and sufficiently fast convergence was observed: saturation was reached for 25 splines per partial wave \cite{KozTup19}. 

It is known, that accurate calculations of the hyperfine structure and parity non-conservation effects for heavy atoms are much more challenging then calculations of the energies, or transition amplitudes.
These properties depend on the wave function at small distances. 
Correlations change the behaviour of the valence-electrons wave functions in this region and lead to the large corrections to the matrix elements of these operators. Because of that the convergence with respect to the number of orbitals in the basis set is typically much slower. 
Test calculations of the magnetic hyperfine constant $A$ for Au did not converge even for the mixed basis set with 45 B-splines per partial wave \cite{KozTup19}. It was found that small components of the valence orbitals near the origin were not smooth. Increasing the size of the basis set decreased the amplitude of these non-physical dips or bumps, but shifted them closer to the origin, where the hyperfine interaction was particularly strong. 
The solution suggested in that paper was to add to the basis set DF orbitals for the ion Au$^{M+}$, or, in other words, the orbitals for the $V^{N-M}$ potential, where N is the number of electrons in the neutral atom.
Such orbitals are similar to those of the neutral atom, but are more contracted. That allowed to smoothly change the orbitals near the nucleus. The calculation of the hyperfine constants $A$ for Au already converged for two such ionic orbitals and 24 B-splines per partial wave. To make this recipe work one need to choose $M$ such that ionic orbitals are neither too different, nor too similar to those of the neutral atom. In the former case such orbitals become less useful, while in the latter one can run into the linear dependency problem. In Ref.\ \cite{KozTup19} the orbitals for the $V^{N-6}$ potential were used.  Here, we suggest another method to form additional orbitals to supplement DF orbitals and B-splines. The new procedure is more formal and does not require arbitrary adjustments. We test these basis sets for calculation of the QED corrections
and hyperfine constants in neutral Cs.

\section{Method}\label{Sec:Method}

The basis set for atomic calculations consists of subsets for different partial waves, which are defined by the relativistic quantum number $\varkappa= (l-j)(2j+1)$, where $l$ and $j$ are the orbital and total angular momenta. The relativistic two-component radial wave functions have the form

\begin{align}\label{eq:DForbital}
    \psi_{n\varkappa}(r) &= \frac{1}{r} 
    \left(\begin{array}{c}
          P_{n\varkappa}(r)  \\ 
          Q_{n\varkappa}(r) 
          \end{array}\right)\,.
\end{align}
Most atomic calculations are done in the no-pair approximation. QED corrections may be included approximately by means of the model operators \cite{FlaGin05,STY13,TuBe13}. This means that the Hamiltonian is projected on the positive energy electronic states and the basis set does not include negative energy continuum. To this end, the small components $Q$ of the virtual basis orbitals can not be chosen arbitrary. Most often they are found from the kinetic balance condition \cite{Stanton_1984}: 
\begin{align}\label{eq:kin_bal}
    Q_{n\varkappa}(r) &= -\frac{\alpha}{2}
    \left(\frac{d}{dr} + \frac{\varkappa}{r}
    \right)P_{n\varkappa}(r)\,, 
\end{align}
where $\alpha\approx \tfrac{1}{137}$ is the fine structure constant (we use atomic units throughout the paper). Below we assume that the small components are always formed with the help of Eq.\ \eqref{eq:kin_bal} and focus on the large components $P$.

When some small perturbation is added to the Hamiltonian, the large component $P$ of the electronic orbital slightly changes. These changes do not affect the asymptotic behaviour at large and small distances and the number of nodes of the function $P$ remains the same. To account for such changes we need to add basic functions with similar properties. Let us consider a simple stretching transformation:   
\begin{align}\label{eq:squeeze}
    P(r) &\to \tilde{P}(r) =P(kr)\,, 
\end{align}
where $k=1+\veps$, $|\veps|\ll 1$. Expanding $\tilde{P}$ in powers of $\veps$ we get:
\begin{align}\label{eq:dif1}
    \tilde{P}(r) &=P(kr)=P\left((1+\veps)r\right)\approx P(r) +\veps r \frac{dP}{dr} \,. 
\end{align}
Thus, if we want to guarantee that stretched DF orbital $\tilde{P}_{n\varkappa}$ can be accurately expanded in the basis set, we can add a virtual orbital: 
\begin{align}\label{eq:ansatz}
    P_{n'\varkappa}(r) &= r \frac{dP_{n\varkappa}}{dr}\,. 
\end{align}
Note that this function has the same asymptotic behavior at $r\to 0$ and at $r\to\infty$ as $P_{n\varkappa}$, however it has an extra node. 
\begin{figure*}[htb]
    \includegraphics[width=1.2\columnwidth]{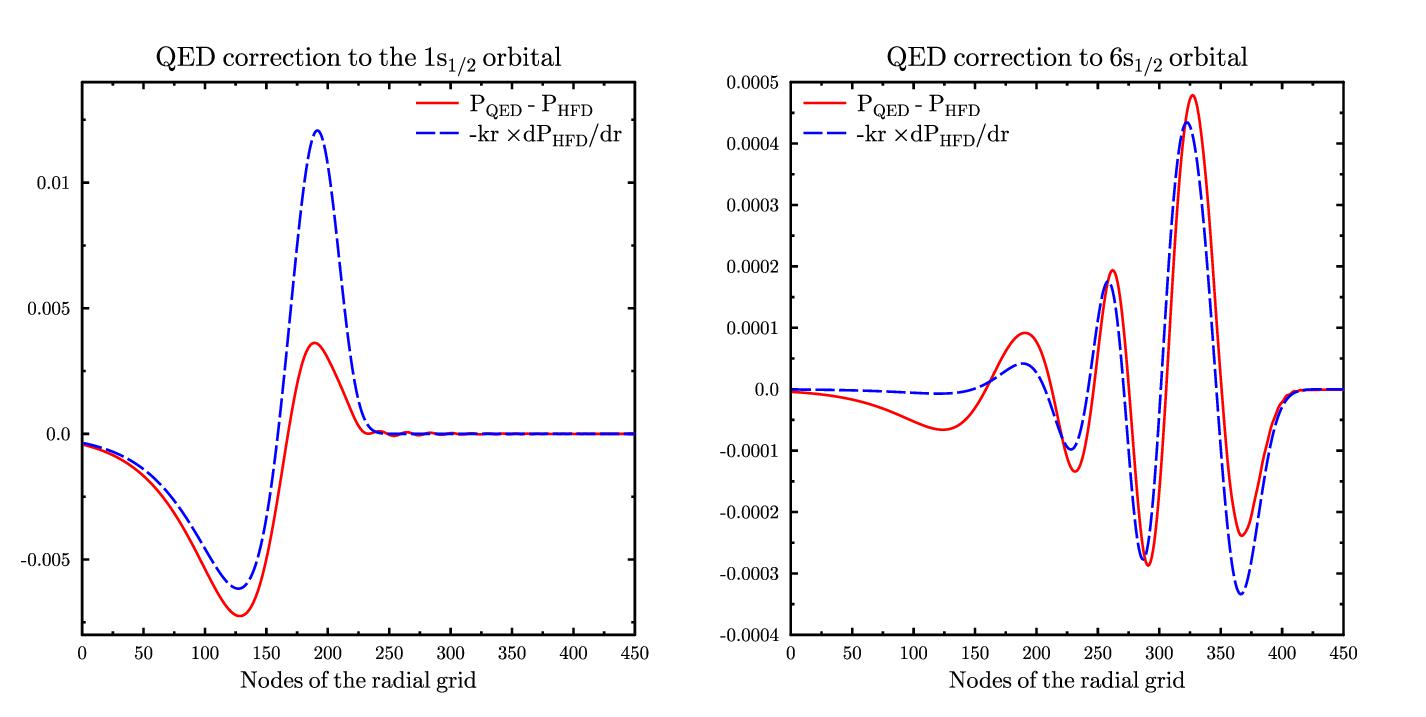}
    \caption{Comparison of the differential ansatz \eqref{eq:ansatz} with the difference between DF orbitals for the Dirac-Coulomb Hamiltonian with and without model QED potential.}
    \label{fig:dif-QED}
\end{figure*}

\begin{table*}[htb] 
    \centering
\caption{ Comparison of the basis sets B7 and B30 for Cs. Results for the QED corrections to the valence binding energies and reduced E1 transition amplitudes and for the RPA corrections to the hyperfine constants $A$. Column three gives DF values. Next two columns give QED or RPA corrections calculated on the basis set B7 with 7 virtual orbitals per partial wave and basis B30 with 30 virtual orbitals per partial wave. Last column gives the difference between two calculations in percent.
\label{tab:QEDs-l}}
\begin{tabular}{lcrrrr}
\toprule
\textbf{Property} & \textbf{Units} & \textbf{DF value}	& \multicolumn{3}{c}{\textbf{QED correction}}\\
&	& & \textbf{Basis B7}	& \textbf{Basis B30}	& \textbf{Diff.}\\
\midrule
$\varepsilon(6s_{1/2})$ & (cm$^{-1}$) & 27954.1 &$-15.798    $&$-15.782    $&$ -0.1\% $\\
$\varepsilon(6p_{1/2})$ &	          & 18790.5 &$  0.814    $&$  0.805    $&$ -1.1\% $\\
$\varepsilon(6p_{3/2})$ &             & 18388.8 &$  0.169    $&$  0.165    $&$ -2.6\% $\\
$\varepsilon(5d_{3/2})$ &             & 14138.5 &$  2.177    $&$  2.151    $&$ -1.2\% $\\
$\varepsilon(5d_{5/2})$ &             & 14162.6 &$  2.063    $&$  2.032    $&$ -1.5\% $\\
\midrule
$E1(6s_{1/2}-6p_{1/2})$	      & (a.u.)  & 5.27769 &$ 0.00322 $&$ 0.00322 $&$  0.0\% $\\
$E1(6s_{1/2}-6p_{3/2})$	      &         & 7.42643 &$ 0.00476 $&$ 0.00475 $&$ -0.2\% $\\
$E1(6p_{1/2}-5d_{3/2})$	      &         & 8.97833 &$-0.00159 $&$-0.00175 $&$ 10.1\% $\\
$E1(6p_{3/2}-5d_{3/2})$	      &         & 4.06246 &$-0.00069 $&$-0.00076 $&$ 10.1\% $\\
$E1(6p_{3/2}-5d_{5/2})$	      &         &12.18640 &$-0.00180 $&$-0.00190 $&$  5.6\% $\\
\midrule
              &       & DF value & \multicolumn{3}{c}{RPA correction}\\
$A(6s_{1/2})$ & (MHz) & 1423.3  &$ 297.6 $&$  293.3  $&$ -1.4\% $\\
$A(6p_{1/2})$ &       & 160.88  &$ 40.02 $&$  40.56  $&$  1.3\% $\\
$A(6p_{3/2})$ &       & 23.916  &$ 19.12 $&$  18.91  $&$ -1.1\% $\\
\bottomrule
\end{tabular}
\end{table*}
\section{Test calculations for Cs}

To illustrate usefulness of the ansatz \eqref{eq:ansatz} let us consider QED corrections in neutral Cs.
Firstly, we used a sufficiently large basis set to diagonalize the Dirac-Fock operator together with the model QED potential of Ref. \cite{STY13,STY15}. After that we calculated the difference between large components with and without QED corrections and compared them with the respective scaled orbitals \eqref{eq:ansatz}. Figure \ref{fig:dif-QED} shows that the latter very much resemble the former. Thus, we can expect that adding orbitals \eqref{eq:ansatz} to the basis set can significantly improve its quality and speeds up the convergence. 

We formed the basis set for Cs which included DF orbitals $1s-7s$, $2p-7p$, and $3d-6d$. For each DF orbital we added respective virtual orbital \eqref{eq:ansatz}. To have the same number of virtual orbitals per partial wave, we formed one additional $p$ orbital per partial wave and three additional $d$ orbitals per partial wave using the method of Ref. \cite{KPF96}. Thus, this basis set included 7 virtual orbitals per partial wave. This calculation was compared with calculation on the long basis set, which included these 7 plus 23 additional virtual orbitals per partial wave.
Below, we refer to these basis sets as B7 and B30, respectively.
We see that the QED correction for the valence $6s$ orbital is practically the same for both basis sets. For the orbital $1s$, both the basis sets produce some non-physical oscillations, but their amplitude for the large basis set is few times smaller.  

\begin{figure*}[htb]
    \centering
    \includegraphics[scale=0.95]{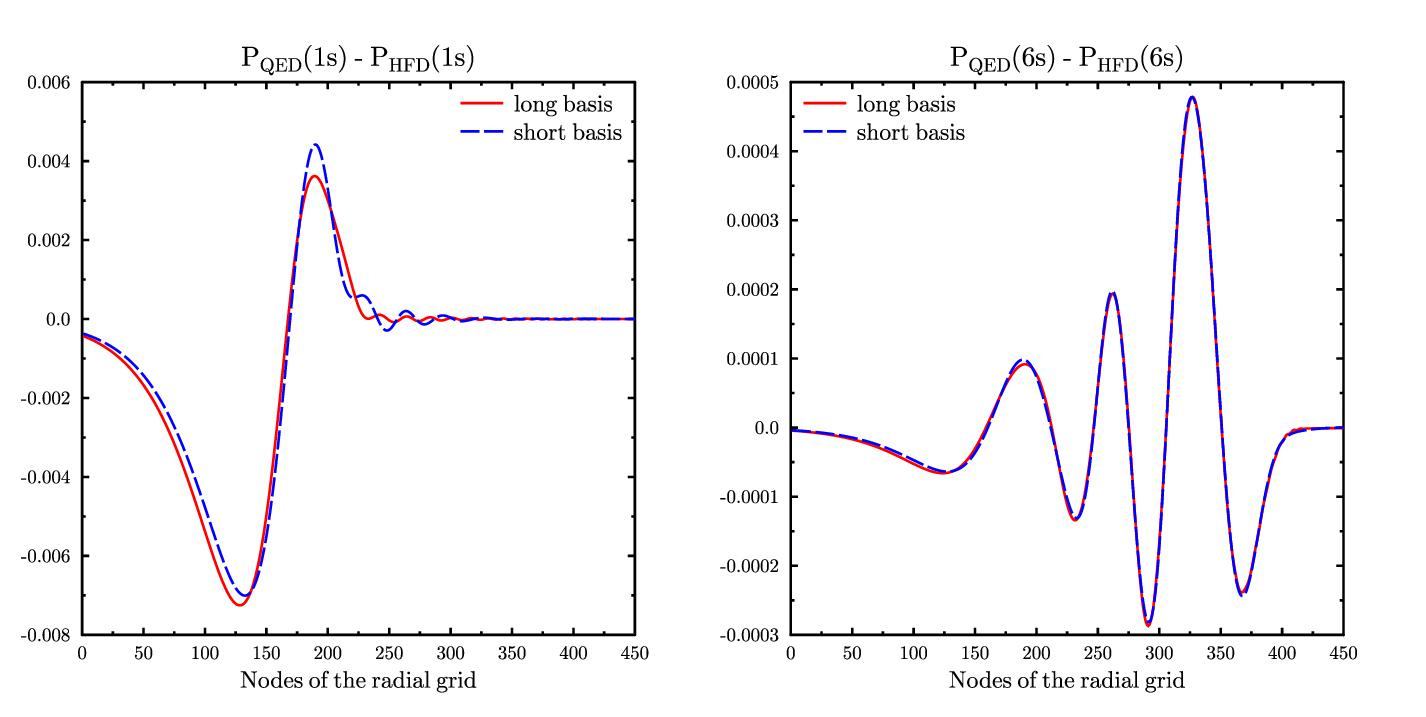}
    \caption{Comparison of the QED calculations with the short (B7) and the long (B30) basis sets. The short basis set includes 7 orbitals per partial wave. In the $s$ wave all of them are formed using ansatz \eqref{eq:ansatz}, while in other partial waves additional orbitals were added using method \cite{KPF96}.}
    \label{fig:QED_short-long}
\end{figure*}

In the calculations of neutral atoms we usually explicitly need only wave functions of the valence electrons, where our short basis set seems to work nicely. To test it further we calculated QED corrections to the valence energies and to the E1 transition amplitudes. In both cases we used model QED potential \cite{STY13}. This model potential accounts for the QED corrections to the wave functions. For E1 amplitudes there are also QED corrections to the vertex, which are not included here. They were calculated in Refs.\ \cite{SaPaChe04,SaChe05} and found to be much smaller. Results obtained with the basis sets B7 and B30 are compared in Table \ref{tab:QEDs-l}. We see that two calculations give practically identical QED correction to the binding energy of the $6s$ electron. For the other partial waves the corrections to the energies are much smaller and the difference between results is about 1~--~2\%. For the $s-p$ transition amplitudes the agreement is again very good, but for $p-d$ transitions the difference between two basis sets is about 10\%. On one hand, these corrections are smaller. On the other hand, in $d$ wave we have only four orbitals formed with the ansatz \eqref{eq:ansatz}, while other three orbitals of the basis set B7 are apparently less useful. We mentioned earlier that convergence in calculations of the hyperfine constants $A$ is often very slow. Because of that we compared random-phase approximation (RPA) corrections for these constants obtained with basis sets B7 and B30. For the orbitals $6s$, $6p_{1/2}$, and $6p_{3/2}$ the difference was around 1\%.  

Results presented in Table \ref{tab:QED-E1} show that the basis sets, which include virtual orbitals \eqref{eq:ansatz} provide fast convergence for calculations of the QED and RPA corrections, which are particularly sensitive to the quality of the basis set. Below we compare QED corrections calculated with these basis sets with calculations of other groups and study potential sources of the differences. We focus primarily on the QED corrections to the $E1$ transition amplitudes.

\subsection{QED corrections to the E1 amplitudes in Cs}

QED corrections to the $E1$ amplitudes in Cs were calculated several times. \citet{SaChe05} calculated $6s_{1/2}-6p_{1/2}$ amplitude within one-determinant approximation, while calculations  \cite{Fairhall2023,TaDe23,Roberts2023} included correlations, but neglected the vertex correction. Traditionally QED corrections to the $E1$ amplitude are expressed in terms of the parameter $R$:
\begin{align} \label{eq:R}
    E1 &= E1_0 \left(1 + \frac{\alpha}{\pi} R\right)\,,
\end{align}
where $E1_0$ is the amplitude without QED correction. There are two contributions to the QED correction $R$:
\begin{align} \label{eq:RPO}
    R &= R_\mathrm{PO}+R_\mathrm{V}\,,
\end{align}
where $R_\mathrm{PO}$ accounts for the correction to the orbitals (perturbed orbitals) and $R_\mathrm{V}$ is the vertex correction. The former is supposed to be larger than the latter \cite{FlaGin05} and it can be approximated by model QED potentials. At present, the vertex correction can be calculated only within the consistent QED approach.

\begin{table*}[htb] 
    \centering
\caption{Comparison of the PO QED corrections $R_\mathrm{PO}$ (see Eq.\ \eqref{eq:RPO}) to the $E1$ transition amplitudes in Cs with other calculations. Our calculations were done using model potentials STY \cite{STY13} and FG \cite{FlaGin05}. \label{tab:QED-E1}}
\begin{tabular}{lD{.}{.}{-1}D{.}{.}{-1}D{.}{.}{4}D{.}{.}{-1}D{.}{.}{-1}}
\toprule
\multicolumn{1}{c}{\textbf{Transition}} &\multicolumn{2}{c}{\textbf{This work}} &\multicolumn{1}{c}{\textbf{\cite{SaChe05}}} &\multicolumn{1}{c}{\textbf{\cite{Fairhall2023}}} &\multicolumn{1}{c}{\textbf{\cite{TaDe23}}}\\
 \multicolumn{1}{c}{Method} &\multicolumn{1}{c}{STY} & \multicolumn{1}{c}{FG} 
    & \multicolumn{1}{c}{\textit{Ab initio}} & \multicolumn{1}{c}{FG} & \multicolumn{1}{c}{FG} \\
\midrule
$R_\mathrm{PO}(6s_{1/2}-6p_{1/2})$	      & 0.262 & 0.267 & 0.326   & 0.272 & 0.26  \\
$R_\mathrm{PO}(6s_{1/2}-6p_{3/2})$	      & 0.275 & 0.280 &         & 0.286 & 0.28  \\
$R_\mathrm{PO}(6s_{1/2}-7p_{1/2})$	      &-2.89  & -2.97 &         &-2.80  &-2.89  \\
$R_\mathrm{PO}(6s_{1/2}-7p_{3/2})$	      &-1.84  & -1.84 &         &-1.76  &-1.84  \\
$R_\mathrm{PO}(7s_{1/2}-6p_{1/2})$	      &-0.442 & -0.452&         &-0.433 &-0.44  \\
$R_\mathrm{PO}(7s_{1/2}-6p_{3/2})$	      &-0.369 & -0.376&         &-0.359 &-0.37  \\
\bottomrule
\end{tabular}
\end{table*}

In Table \ref{tab:QED-E1} we compare our results for $R_\mathrm{PO}$ with results of other groups. \citet{SaChe05} made \textit{ab initio} QED calculation using local Kohn-Sham potential. They calculated both corrections and found, that the vertex correction $R_\mathrm{V}$ was significantly smaller, $R_\mathrm{V}=-0.065$. Other calculations used non-local DF operator and model potential \cite{FlaGin05,GiBe16,2016_GingesJ_PhysRevA}. 
Here we made calculations with model potentials \cite{STY13} and \cite{FlaGin05}, which we denote as STY and FG potentials respectively.

Table \ref{tab:QED-E1} shows that the results obtained with two QED potentials agree within 2~--~3\%. The difference between different groups is somewhat bigger, about 4~--~6\%. There are small differences in the model potentials described in Refs.\ \cite{FlaGin05} and \cite{GiBe16,2016_GingesJ_PhysRevA}. Incompleteness of the basis sets can be another reasons of the disagreements. The only direct QED calculation is about 20\% larger, than all three calculations with model potentials. Note that this calculation was done with the local Kohn-Sham potential instead of the DF potential used by all other groups. Because of that, it is important to check, how much QED corrections depend on the choice of the potential. Note also that parametrization \eqref{eq:R} allows to partly compensate for the differences in the value of the initial amplitude $E1_0$ in different approximations.

\subsection{Local screening potentials}

We made calculations with the model QED operator \cite{STY13} and three local screening potentials: the first one is the core-Hartee potential~(CH) induced by the core electrons of Cs$^+$ ground-state configuration, the other two are Kohn-Sham~(KS)~\cite{1965_KohnW_PR140} and Slater~(S)~\cite{1972_SlaterJ_PRB5} potentials corresponding to the ground-state configuration of the neutral Cs.
The proper asymptotic behavior of the potentials KS and S is restored by including the Latter correction \cite{1955_LatterR_PR99}.  
In Table \ref{tab:QED-local} the results obtained using the local screening potentials are compared with the Dirac-Fock results.
We see that calculations of QED corrections $\delta\veps$ with local potentials do not correlate with the DF calculations for all partial waves, but the $s$ wave. The reason for this discrepancy is obvious. In the DF approximation the energy correction comes from the direct contribution of the perturbation and the contribution from the adjustment of the self-consistent field (relaxation of the core) $\delta V_\mathrm{SCF}$. The former is large for the $s$ wave and rapidly decreases with the orbital quantum number $l$. At the same time, the relaxation of the core affects all valence orbitals in a similar way. Two contributions become comparable already for the valence $p$ orbitals. For the local potentials there is only direct contribution. The KS potential reproduces valence orbital energies $\veps$ better than two other potentials, while QED corrections $\delta\veps$ to the $s$ wave energies for the CH and KS potentials are of the similar accuracy. Slater potential is clearly the least accurate.

\begin{table*}[htb] 
    \centering
\caption{QED corrections $\delta\veps$ to the binding energies $\veps$ in Cs calculated in DF approximation and with three local potentials, core-Hartree (CH), Kohn-Sham (KS), and Slater (S). All calculations were done using model QED potential \cite{STY13} (STY).\label{tab:QED-local}}
\begin{tabular}{lD{.}{.}{5}D{.}{.}{2}D{.}{.}{5}D{.}{.}{2}D{.}{.}{5}D{.}{.}{2}D{.}{.}{5}D{.}{.}{2}}
\toprule
\multicolumn{1}{c}{\textbf{Orbital}} 
  &\multicolumn{2}{c}{\textbf{Dirac-Fock}} 
    &\multicolumn{2}{c}{\textbf{core-Hartree}} 
      &\multicolumn{2}{c}{\textbf{Kohn-Sham}} 
        &\multicolumn{2}{c}{\textbf{Slater}}\\
 &\multicolumn{1}{c}{$\veps$} & \multicolumn{1}{c}{$\delta\veps$} 
 &\multicolumn{1}{c}{$\veps$} & \multicolumn{1}{c}{$\delta\veps$} 
 &\multicolumn{1}{c}{$\veps$} & \multicolumn{1}{c}{$\delta\veps$} 
 &\multicolumn{1}{c}{$\veps$} & \multicolumn{1}{c}{$\delta\veps$} 
    \\
 &\multicolumn{1}{c}{(a.u.)} & \multicolumn{1}{c}{(cm$^{-1}$)} 
 &\multicolumn{1}{c}{(a.u.)} & \multicolumn{1}{c}{(cm$^{-1}$)} 
 &\multicolumn{1}{c}{(a.u.)} & \multicolumn{1}{c}{(cm$^{-1}$)} 
 &\multicolumn{1}{c}{(a.u.)} & \multicolumn{1}{c}{(cm$^{-1}$)} 
   \\
\midrule
$6s_{1/2}$ & 0.12737 &-15.80 & 0.11998 &-16.13 & 0.12394 &-16.22 & 0.13573 &-23.54 \\
$6p_{1/2}$ & 0.08562 & 0.80  & 0.08130 &-0.20  & 0.08489 & -0.17 & 0.08908 & -0.20 \\
$6p_{3/2}$ & 0.08379 & 0.16  & 0.07876 &-1.09  & 0.08278 & -0.91 & 0.08676 & -0.96 \\
$5d_{3/2}$ & 0.06442 & 2.15  & 0.05898 & 0.04  & 0.06041 &  0.07 & 0.07735 &  0.38 \\
$5d_{5/2}$ & 0.06453 & 2.03  & 0.05886 &-0.05  & 0.06011 & -0.09 & 0.07570 & -0.49 \\
$7s_{1/2}$ & 0.05519 &-4.30  & 0.05327 &-4.71  & 0.05457 & -4.85 & 0.05791 & -6.40 \\
$7p_{1/2}$ & 0.04202 & 0.29  & 0.04060 &-0.07  & 0.04184 & -0.06 & 0.04334 & -0.07 \\
$7p_{3/2}$ & 0.04137 & 0.05  & 0.03971 &-0.39  & 0.04108 & -0.32 & 0.04251 & -0.34 \\
$6d_{3/2}$ & 0.03609 & 1.22  & 0.03318 & 0.11  & 0.03422 &  0.19 & 0.04152 &  0.24 \\
$6d_{5/2}$ & 0.03609 & 0.88  & 0.03311 &-0.13  & 0.03402 & -0.22 & 0.04106 & -0.32 \\
\bottomrule
\end{tabular}
\end{table*}

Taking into account that core relaxation is very important for the $p$ and $d$ waves, the local approximation is clearly inapplicable for the QED corrections to the $E1$ transitions between $p$ and $d$ orbitals. The results of the calculations for the $s \leftrightarrow p$ transitions are presented in Table \ref{tab:QED-local-E1}. Again, the KS potential gives better agreement with DF, than two other potentials. In most cases the difference is less, or about 10\%. The only exception is the $6s_{1/2}-7p_{1/2}$ transition, where the difference is 24\%. For the CH and S potentials the average differences are about 14\% and 22\% respectively.

\begin{table*}[htb] 
    \centering
\caption{Comparison of the PO QED corrections $R_\mathrm{PO}$ to the $s \leftrightarrow p$ transition amplitudes in Cs for different potentials. For local potentials we also give the difference with DF results in percent. All calculations were done using model QED potential \cite{STY13} (STY).\label{tab:QED-local-E1}}
\begin{tabular}{lD{.}{.}{5}D{.}{.}{5}rD{.}{.}{5}rD{.}{.}{5}r}
\toprule
\multicolumn{1}{c}{\textbf{Transition}} 
  &\multicolumn{1}{c}{\textbf{Dirac-Fock}} 
    &\multicolumn{2}{c}{\textbf{core-Hartree}} 
      &\multicolumn{2}{c}{\textbf{Kohn-Sham}} 
        &\multicolumn{2}{c}{\textbf{Slater}}\\
\midrule
$R_\mathrm{PO}(6s_{1/2}-6p_{1/2})$  & 0.266 & 0.299 & 12\%  & 0.281  &  6\% & 0.365 & 37\%   \\
$R_\mathrm{PO}(6s_{1/2}-6p_{3/2})$  & 0.274 & 0.316 & 15\%  & 0.296  &  8\% & 0.389 & 42\%   \\
$R_\mathrm{PO}(6s_{1/2}-7p_{1/2})$  &-2.856 &-2.646 &  7\%  &-3.541  & 24\% &-3.000 &  5\%   \\
$R_\mathrm{PO}(6s_{1/2}-7p_{3/2})$  &-1.810 &-1.388 & 23\%  &-1.945  &  7\% &-1.814 &  0\%   \\
$R_\mathrm{PO}(7s_{1/2}-6p_{1/2})$  &-0.446 &-0.417 &  7\%  &-0.433  &  3\% &-0.492 & 10\%   \\
$R_\mathrm{PO}(7s_{1/2}-6p_{3/2})$  &-0.373 &-0.311 & 17\%  &-0.341  &  8\% &-0.403 &  8\%   \\
$R_\mathrm{PO}(7s_{1/2}-7p_{1/2})$  & 0.255 & 0.311 & 22\%  & 0.284  & 11\% & 0.372 & 46\%   \\
$R_\mathrm{PO}(7s_{1/2}-7p_{3/2})$  & 0.292 & 0.318 &  9\%  & 0.289  &  1\% & 0.380 & 30\%   \\
\bottomrule
\end{tabular}
\end{table*}

We conclude that for Cs the KS potential provides higher accuracy for the QED calculations than CH and S potentials.  
Note, that the result of \citet{SaChe05} for the $6s_{1/2}-6p_{1/2}$ transition cited in Table \ref{tab:QED-E1} was obtained using KS potential. Our KS value is closer to their value, then the DF one. Thus, the approximately 20\% difference between \textit{ab initio} calculation \cite{SaChe05} and calculations with model QED potentials presented in Table \ref{tab:QED-E1} can be attributed in part to the local potential error.

\section{Conclusions}

Many numerical methods for atomic calculations require one-electron basis sets. Here we suggested the differential ansatz to generate virtual orbitals from the DF orbitals of the core and valence electrons. The number of such orbitals cannot exceed the number of DF orbitals. More virtual orbitals can be added using B-splines, or other traditional methods. Our test calculations for Cs show good convergence of calculations of different properties, including QED corrections and hyperfine structure constants. 

We also tested three local potentials, which are often used for the QED calculations in many-electron atoms and found out that for the lower partial waves the Kohn-Sham local potential demonstrated best agreement with the DF potential. For the higher
partial waves the results obtained with all tested local potentials are very different from the results obtained with the non-local DF potential. These differences are caused by the absences of the contribution from the change of the self-consistent field induced by the perturbation. This effect becomes dominant for the partial waves with $l>1$ making calculation of the respective QED corrections problematic.
In order to improve the accuracy of the \textit{ab initio} QED calculations in heavy atoms it is necessary to
include core-relaxation effects by going to the next order of the perturbation theory in the electron-electron interaction. 
The next step in quantum-mechanical calculations of the QED corrections requires to construct effective operators for the vertex contribution.

\acknowledgments{We are grateful to Ilya Tupitsyn and Vladimir Yerokhin for extremely useful discussions and suggestions. This work was supported by the Russian Science Foundation grant \# 23-22-00079.}
\bibliographystyle{apsrev}

\end{document}